# Simulating SiD Calorimetry: Software Calibration Procedures and Jet Energy Resolution

Ron Cassell[1]

1 – SLAC National Accelerator Laboratory - USA

Simulated calorimeter performance in the SiD detector is examined. The software calibration procedures are described, as well as the perfect pattern recognition PFA reconstruction. Performance of the SiD calorimeters is summarized with jet energy resolutions from calorimetry only, perfect pattern recognition and the SiD PFA algorithm. Presented at LCWS08[1].

## 1 Introduction

Our objective is to simulate the calorimeter performance of the SiD detector, with and without a Particle Flow Algorithm (PFA). Full Geant4 simulations using SLIC[2] and the SiD simplified detector geometry (SiD02) are used. In this geometry, the calorimeters are represented as layered cylinders. The EM calorimeter is Si/W, with 20 layers of 2.5mm W and 10 layers of 5mm W, segmented in 3.5x3.5mm$^2$ cells. The HAD calorimeter is RPC/Fe, with 40 layers of 20mm Fe and a digital readout, segmented in 10x10mm$^2$ cells. The barrel detectors are layered in radius, while the endcap detectors are layered in z(along the beam axis).

The software calibration is a means of converting the simulated detector response into an energy measurement. While this would be straightforward in a single calorimeter with a perfectly linear response, there are several problems with combining multiple systems each with a different response. The responses vary with particle type (photon, charged hadron, neutral hadron), energy (with many sources of nonlinearity for hadrons), polar angle, depth of interaction, and even type of neutral hadron (n,nbar, $K^0_L$). In principal a detailed study of each dependence for each calorimeter is possible, but in practice simplified procedures averaging over most of the effects were developed.

In a PFA, individual particles are reconstructed. Therefore separate calibrations were performed for photons and hadrons, allowing a different conversion of detector response to energy, depending on particle type. Since the objective is to optimize jet energy resolution in physics events, we used a representative physics process, ZZ events at 500 GeV, to provide a reasonable average energy spectra and particle mix. To characterize the performance of SiD purely as a calorimeter, another calibration was performed, averaging over all dependencies with no assumption as to the source of the energy deposits.

## 2 Photon Calibration and Performance

For the photon calibration, only the response of the EM calorimeters was considered. Photons and their associated EM calorimeter hits from the ZZ dataset were selected, with the requirement E > 1GeV and cosθ < 0.9. A "sampling fraction" ($S_F$) is defined as a single conversion factor from energy measured in a sampling calorimeter to actual energy, such that $E_{Meas} = E_{Deposited}/S_F$. A $S_F$ was calculated for each calorimeter (4 parameters) by minimizing the sum over all photons of $(dE/\sqrt{E_{Gen}})^2$, where $dE = E_{Meas} - E_{Gen}$. The four parameters arise from treating the barrel and endcap calorimeters separately, and the two different absorber thicknesses in each EM calorimeter. Many effects contribute to differences in the barrel and endcap calorimeters, i.e. incident angle of the photons, angular extent of cell segmentation and magnetic field orientation. Since we are averaging over many effects, separate sampling fractions were calculated for the barrel and endcap calorimeters. The assumption of a linear response is implicit in the use of a single sampling fraction per calorimeter. The photon energy distribution from this dataset is shown in Fig. 1, and the result of this photon calibration is shown in Fig. 2 as an "effective" resolution of $19\%/\sqrt{E_{Gen}}$ for all photons in the sample. The results were then applied to single photons from ttbar events, with linearity and resolution curves shown in Figs. 3 and 4. The maximum nonlinearities are $\sim 0.5\%$. While many improvements are possible, the results were deemed adequate for jet reconstruction with a PFA, and no further corrections were applied.

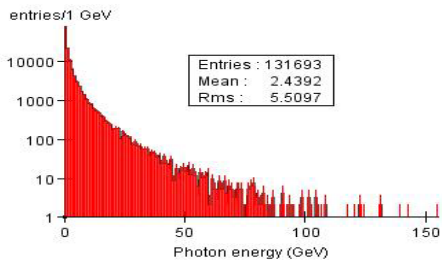

**Figure 1:** Single photon generated energy distribution in ZZ events at $\sqrt{s} = 500$ GeV.

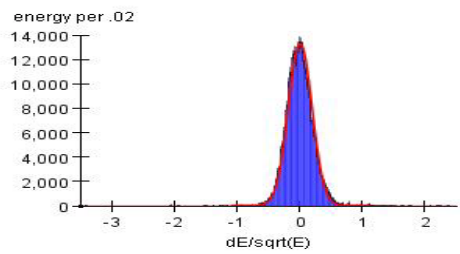

**Figure 2:** Energy weighted $dE/\sqrt{E_{Gen}}$ distribution for single photons in the ZZ dataset. For the Gaussian fit, sigma = 0.188.

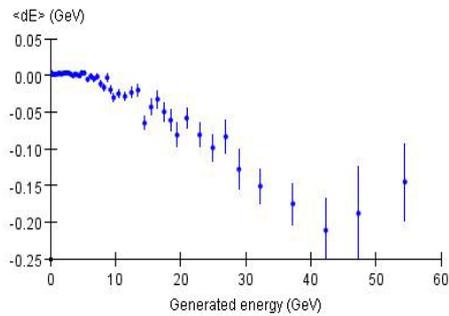

**Figure 3:** Linearity plot for single photons from ttbar events. The mean value of the residuals is plotted vs

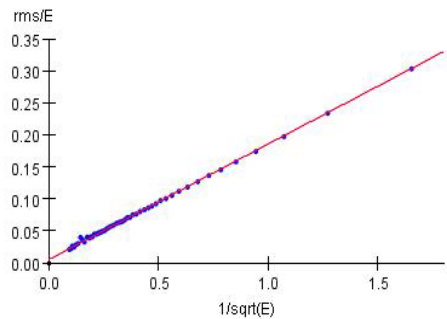

**Figure 4:** Resolution plot for single photons from ttbar events. The line in the resolution plot represents a fit to the data, $0.5\% + 18\%/\sqrt{E_{Gen}}$.

## 3  Neutral Hadron Calibration and Performance

For the neutral hadron calibration, the response of the (4) EM calorimeters and (2) HAD calorimeters (barrel and endcap) was considered. Prompt neutral hadrons and their associated calorimeter hits from the ZZ dataset were selected, with the requirement E > 2GeV, cosθ <

0.9 and no interaction occurred before reaching the EM calorimeter. In studies of the isolated digital RPC/Fe HAD calorimeter[3], a correction of the form 1+a(cscθ-1), where a=−0.23, was found to represent the data adequately. This correction was applied to the hits in the HAD calorimeters before finding sampling fractions. A sampling fraction per calorimeter was calculated minimizing the sum of $(dE/\sqrt{E_{Gen}})^2$ over all neutral hadrons. There are many sources of non-linearity for neutral hadrons. They include leakage, saturation, response to different particle types and Geant4 modeling. Although the implicit assumption of linearity in using simple sampling fractions is far from true for neutral hadrons, these sampling fractions were used to add contributions from different calorimeters for a "combined" energy($E_{SF}$), and corrections applied to that sum. The sampling fractions obtained (compared with the photon sampling fractions) are shown in Table 1. The neutral hadrons were then separated in bins of $E_{SF}$, and a set of points {<$E_{SF}$>,<$E_{Gen}$>} was obtained. By linearly interpolating between these points, the observed $E_{SF}$ is converted to the energy measurement. The input energy distribution and the effective resolution (63%/$\sqrt{E_{Gen}}$) for this sample are shown in Figs. 5 and 6. Neutral hadrons from ttbar events at 500 GeV were used to check the results, shown in Figs. 7 and 8. The first plot shows the mean value of the residuals for both $E_{SF}$ and the corrected energy measurement. The abrupt change between 10 and 15 GeV is due to Geant4 modeling. After corrections, the nonlinearities are ~3%. Additional corrections have been attempted, yielding marginal improvements in neutral hadron resolution and no significant improvement in jet energy resolution, so this simplified software calibration was used for reconstruction.

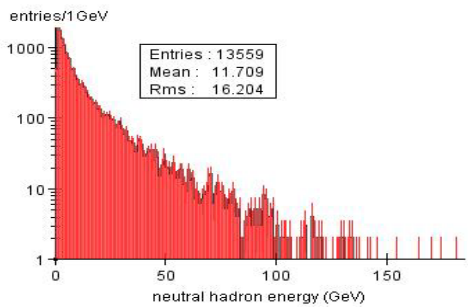

**Figure 5:** Single neutral hadron generated energy distribution in ZZ events at √s = 500 GeV.

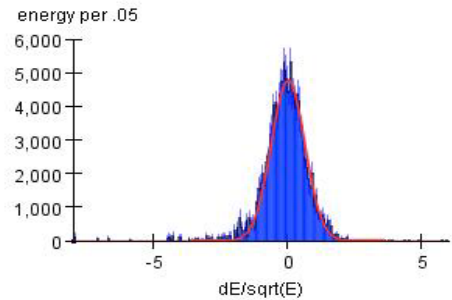

**Figure 6:** Energy weighted dE/$\sqrt{E_{Gen}}$ distribution for single neutral hadrons in the ZZ dataset. For the Gaussian fit, sigma = 0.625.

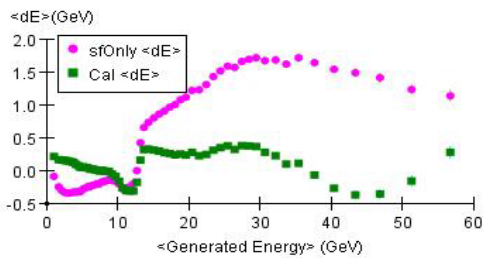

**Figure 7:** Linearity plot for single neutral hadrons from ttbar events. The mean value of the residuals is plotted vs the generated energy using only the

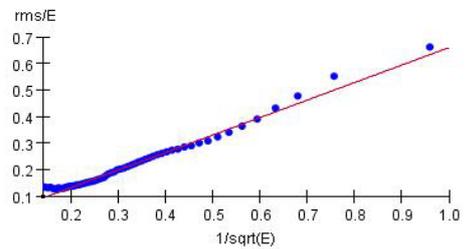

**Figure 8:** Resolution plot for single neutral hadrons from ttbar events. The line in the resolution plot represents 65%/$\sqrt{E_{Gen}}$.

| Calorimeter | EMBarrel Inner | EMBarrel Outer | EMEndcap Inner | EMEndcap Outer | HADBarrel | HADEndcap |
|---|---|---|---|---|---|---|
| Photon $S_F$ | 0.0180 | 0.00890 | 0.0172 | 0.00813 | | |
| NHad $S_F$ | 0.0149 | 0.00855 | 0.0149 | 0.00853 | 9.41 0.106 GeV/hit | 8.43 0.119 GeV/hit |

**Table 1:** Values of the sampling fractions calculated for each calorimeter from photon and neutral hadron calibrations. The split EM calorimeters reflect the change

## 4 Charged Hadron Calibration

In a PFA, the measurements from the tracking systems are used to reconstruct the charged particle momenta. However, the assignment of the calorimeter hits to the charged tracks is a critical part of the algorithm. Some form of E/p matching is used to reduce pattern recognition mistakes, therefore a conversion of calorimeter hits to energy is needed for charged hadrons. In Geant4 studies of sampling calorimeters[3], it was shown that the response was equivalent to that of neutral hadrons if one considered the pre-interaction mip trace separately. Therefore the conversion of calorimeter hits to energy for charged hadrons is done by calculating the mean dE/dx for the initial mip trace, and the neutral hadron calibration is used for the remaining hits.

## 5 Pure Calorimetric Performance

As a check of calorimeter performance, a sampling fraction per calorimeter (8 parameters, muon system included) was calculated with no assumption as to type of particle depositing energy. To obtain the sampling fractions, qqbar events (q = uds) at fixed energies were used. With no radiation and no prompt neutrinos the total energy per event was fixed. Equal number of events from CM energy = (100,200,360,500) GeV were used. For each event, the energy deposits in each calorimeter were summed. For the analog EM calorimeters, no angle correction was applied. The hits in the HAD and muon calorimeters were corrected for polar angle using the form described in Section 3. The sampling fractions were then obtained from a fit minimizing the sum of $(dE/\sqrt{E_{Gen}})^2$ over all events. Using these sampling fractions, the results of a purely calorimetric energy measurement are shown in Fig. 9. The expression rms90/E = alpha90/$\sqrt{E}$ defines alpha90, where rms90 is the rms of the 90% of the events yielding the smallest rms. Until leakage and/or saturation becomes significant, SiD behaves as

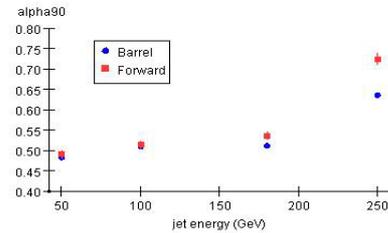

**Figure 9:** α90 vs jet energy for

~50%/√E calorimetric detector. (Using rms90 as the resolution).

## 6 Jet Energy Resolution with Perfect Pattern Recognition

For the perfect pattern recognition reconstruction, a cheat tracking program was used. Charged Monte Carlo particles with a sufficient number of hits were called "trackable", and the track parameters were smeared. For each event a set of final state particles was defined, starting with the generator final state particles and replacing those particles that interacted or decayed before entering the calorimeters with the interaction/decay products if a "trackable" particle was produced.

For each of the final state particles a cheat reconstruction was performed. For "trackable" charged particles, the smeared track parameters were used to create a reconstructed particle, and the mass assigned as a pion mass, except for electrons and muons where the appropriate mass was used. For all other particles the Monte Carlo truth information was used to assign calorimeter hits to the particles. Each particle with sufficient number of calorimeter hits was used to create a neutral reconstructed particle, with the previously described calibration used to determine energy and the mean hit position used to determine direction. The mass was assigned 0 for photons and the $K^0_L$ mass for all others.

This reconstruction was performed on the qqbar and ZZ test samples. The qqbar samples contain two uds quark jets of equal energy with no prompt neutrinos. This allows measurement of the jet energy resolution without a jet finding algorithm by summing the energy of all reconstructed particles in the event. In the ZZ test sample, one Z decays to two neutrinos and the other Z decays to two uds quarks. This allows measurement of the dijet mass resolution, again without jet finding, by summing the energy and momentum of all reconstructed particles in the event. The results of this reconstruction on the qqbar and ZZ test samples are shown in Figs. 10 and 11. Using rms90 as the measure of resolution, jet energy resolutions are between 17%/√E and 24%√E, and the dijet mass resolution dM/M ~ 2.5%.

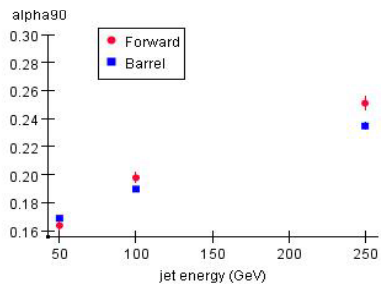

**Figure 10:** α90 vs jet energy for perfect pattern recognition reconstruction.

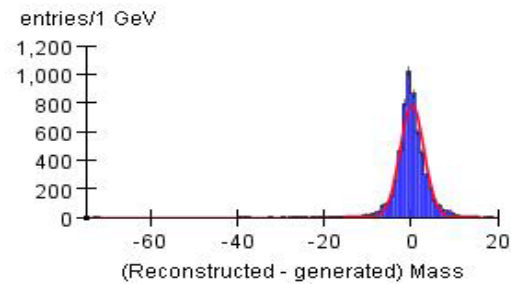

**Figure 11:** Mass residuals in ZZ events using the perfect pattern recognition reconstruction. rms90 = 2.24 GeV, and for the Gaussian fit σ = 2.73 GeV.

## 7 Jet Energy Resolution with the SiD PFA

The SiD PFA is described elsewhere in these proceedings [4,5]. The jet energy resolution is compared to the perfect pattern recognition result in Figure 12. Both use the same events with the same calibration. The jet energy resolution from the SiD PFA varies from ~35%/√E to 70%/√E, as the jet energy varies from 50-250 GeV. This translates to dE/E for these jets of 3.75%-5.25%. The mass residuals (reconstructed − generated) from ZZ events using the SiD PFA are shown in Fig. 13, yielding dM/M ~ 4.5%. Algorithm development continues, with the emphasis on better pattern recognition.

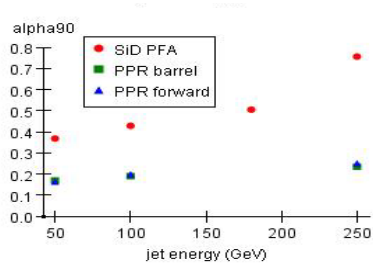

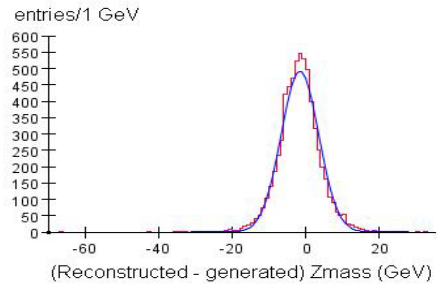

**Figure 12:** α90 vs jet energy for the SiD PFA and the perfect pattern recognition reconstruction.

**Figure 13:** Mass residuals in ZZ events using the SiD PFA reconstruction. rms90 = 4.00 GeV, and the Gaussian fit σ = 5.11 GeV.

## 8   Acknowledgements

Work supported by the U.S. Department of Energy under contract number DE-AC02-76SF00515.